\documentclass[showpacs,twocolumn,prl,aps]{revtex4}
\usepackage{graphicx}

\begin{document}

\title[Short title for running header]{Topological solid phase in a quantum dimer model}
\author{Jianhua Yang and  Tao Li}
\affiliation{ Department of Physics, Renmin University of China,
Beijing 100872, P.R.China}
\date{\today}

\begin{abstract}
We present an example for the phase transition between a topological non-trivial solid phase and a trivial solid phase in the quantum dimer model(QDM) on triangular lattice. Such a transition is beyond the Landau's paradigm of phase transition. We have characterized such a transition with the topological entanglement entropy(TEE) of the system, which is found to change from $\gamma=\ln 2$ in the topological solid phase to zero in the trivial solid phase, through a pronounced peak around the transition point. We also calculated the correlation function of the vison excitation in the QDM and find that the vison condensate develops right at the topological transition point. These results imply that the topological order and the related fractionalized excitation can coexist with conventional symmetry breaking order.
\end{abstract}
 \maketitle

Topological order is a novel form of order that goes beyond the Landau's paradigm of phase transition\cite{Wen1}. The topological order manifests itself in both the topological degeneracy for system defined on multiply connected manifold and the topological  contribution to the entanglement entropy. The concept of topological order plays an important role in the study of fractional quantum Hall systems\cite{Wen2}, the spin liquid phase in quantum magnets\cite{Read,Wen3} and the pseudogap phase in the high-T$_{c}$ superconductors\cite{Senthil,Scheurer}. A direct physical consequence of the topological order is the emergence of fractionalized excitation in the spectrum of the system.

The concept of topological order is best illustrated by the quantum dimer model(QDM)\cite{RK,Sondhi}, whose degree of freedom is the dimer occupying the nearest neighboring bonds of the lattice. The dimer in the QDM can be viewed as a simplified version of the spin singlet pair in the resonating valence bond theory of quantum spin liquid\cite{Anderson} and should satisfy the constraint that each site of the lattice should be occupied by one and only one dimer. On multiply connected manifold, the Hilbert space of the QDM factorizes into distinct topological sectors that can not be connected by any local Hamiltonian term at finite order. For example, on a two dimensional torus, we can classify the dimer configurations into four topological sectors according to the parity of the number of dimers that are intersected by two large loops wrapping around the two holes of the torus. The $Z_{2}$ topological degeneracy of the QDM defined on the torus results from the fact that different topological sectors are indistinguishable by local Hamiltonian. 

The local order parameter related to the conventional symmetry breaking order can in principle be used to distinguish dimer configuration in the different topological sectors of the QDM. It is thus unclear if the topological order in the QDM is robust or not when one introduce symmetry breaking order in the QDM. For example, the absence of the $Z_{2}$ topological order can be easily seen when the dimers in the QDM froze into a static valence bond solid dimer configuration of either the staggered or columnar type, which is a simple product state. Spinon excitation is obviously confined in such a classical background. More generally, it is widely believed that the topological order and the related fractionalized excitation should be found in system without any conventional symmetry breaking order. However, a fully symmetric state is usually very difficult to realize in the strict sense. On the other hand, spectral signature that is akin to the continuum of fractionalized excitation has been reported in more than one quantum magnetic systems with long-range magnetic order\cite{AF1,AF2,AF3}. It is thus crucial to know if the conventional symmetry breaking order can coexist with topological order. 

The purpose of this paper is to provide a concrete example for the coexistence of the topological order and the conventional symmetry breaking order in a QDM. The existence of the topological order is here proved by the non-trivial topological entanglement entropy(TEE) of the system. We also show that the transition between the topological solid phase and the trivial solid phase is driven by the condensation of the vison excitation.

\begin{figure}[h!]
\includegraphics[width=8cm,angle=0]{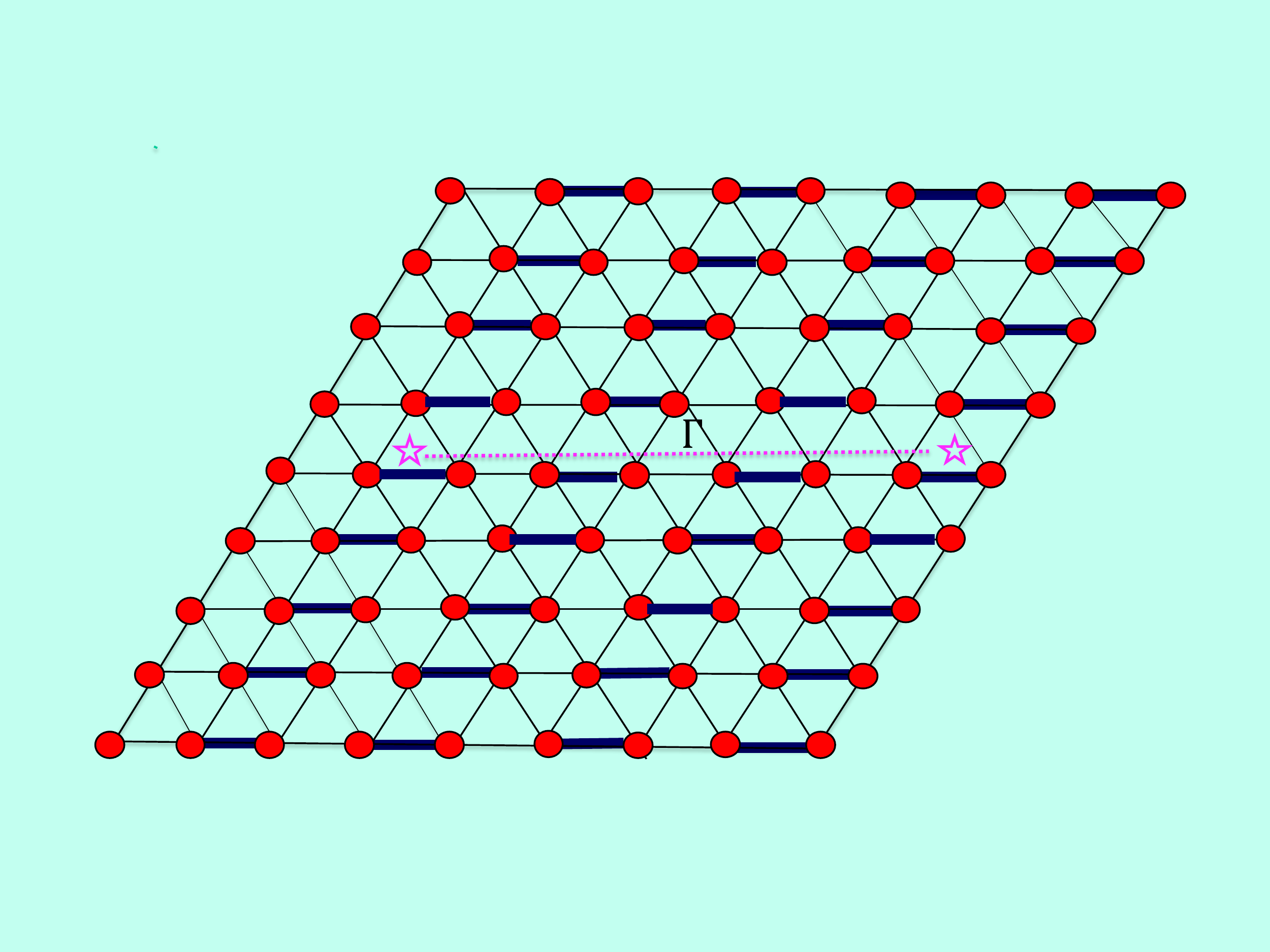}
\caption{Illustration of the generalized R-K state of the triangular QDM studied in this work. The system is defined on a $L\times L$ torus. Each dimer occupying the thick bonds contribute a factor of $\eta$ to the wave function of the system. The two pink stars denote the source of two vison excitations, which are connected by the cut line $\Gamma$. The correlation function between the two visons is given by the expectation value of $(-1)^{n_{\Gamma}}$, in which $n_{\Gamma}$ is the number of dimers that are intersected by $\Gamma$.} \label{fig1}
\end{figure}

The quantum state used to demonstrate the coexistence of topological order and conventional symmetry breaking order in the QDM is given by the following generalized Rokhsar-Kivelson(R-K) wave function 
\begin{equation}
|\Psi\rangle=\sum_{C}\psi(C)|C\rangle.\nonumber
\end{equation}
Here $|C\rangle$ denotes a general dimer covering of the triangular lattice. The amplitude $\psi(C)$ is given by $\psi(C)=\eta^{n_{t}}$, in which $n_{t}$ is the number of dimers that occupy the thick bonds illustrated in Figure 1. $|\Psi\rangle$ reduces to the standard R-K wave function of the triangular QDM when $\eta=1$, which is a well known example of $Z_{2}$ topological liquid\cite{Sondhi,Fendley,Ioselevich}. When $\eta> 1$, the lattice symmetry is broken explicitly to establish a columnar valence bond solid order\cite{Sondhi,Ralko1}. It is not clear if the topological degeneracy is robust or not in this situation. 

To diagnose the topological characteristic of the system, we have calculated the TEE of the system as a function of $\eta$. The topological entanglement entropy $\gamma$ is defined as the universal part of the correction to the entanglement entropy from the area law\cite{Kitaev,Levin}, namely, $S=\alpha A- \gamma$. Here $A$ is the area of the boundary between the subsystem and its complement, $\alpha$ is a non-universal constant depending on the details of the boundary. For numerical convenience, we will consider the Renyi entanglement entropy of the second order, which is given by $S_{2}=-\mathrm{ln}\mathrm{Tr}\rho_{s}^{2}$. Here $\rho_{s}$ is the reduced density matrix of the subsystem and is given by $\rho_{s}=\mathrm{Tr}_{e} | \Psi \rangle \langle \Psi |$. Here the trace $\mathrm{Tr}_{e}$ is over the degree of freedoms in the environment. We note that the value of $\gamma$ is independent of the order of the Renyi entanglement entropy\cite{Renyi}.

The trace of $\rho_{s}^{2}$ involved in $S_{2}$ can be calculated with the replica trick\cite{Hastings}, in which one introduces an identical copy of the system as its replica and divide it into subsystem and environment in exactly the same manner as for the system. It can be shown that
\begin{eqnarray}
\mathrm{Tr}\rho_{s}^{2}=\langle\Psi \otimes \Psi
|\widehat{\mathrm{SWAP}}|\Psi \otimes \Psi\rangle,\nonumber
\end{eqnarray}
in which $|\Psi \otimes \Psi\rangle$ denotes the direct product of the state vector of the system and its replica, $\widehat{\mathrm{SWAP}}$ denotes the operation of
exchanging all the degree of freedoms within the subsystem between
the system and its replica\cite{Hastings}. With such a trick, $\mathrm{Tr}\rho_{s}^{2}$ can be represented as
\begin{eqnarray}
\mathrm{Tr}\rho_{s}^{2}=\frac{\sum_{C_{1},C_{2}}S(C_{1},C_{2})}{\sum_{C_{1},C_{2}}}.
\end{eqnarray}
Here $C_{1}$ and $C_{2}$ denote dimer configuration in the
system and the replica respectively. $S(C_{1},C_{2})$ is $1$ if the swapped configuration satisfy the dimer
constraint and is otherwise zero.

In a previous work, a very efficient ratio trick is proposed to evaluate the above trace with the Monte Carlo method\cite{Pei}. More specifically, $\mathrm{Tr}\rho^{2}_{s}$ can be represented as the product over a series of ratios as follows
\begin{eqnarray}
\mathrm{Tr}\rho_{s}^{2}=\prod_{i=1,m}\frac{\sum_{C_{1},C_{2}}(r_{i-1})^{N_{err}}}{\sum_{C_{1},C_{2}}(r_{i})^{N_{err}}},\nonumber
\end{eqnarray}
in which $r_{i}\in [0,1]$ is a series of real numbers satisfying $r_{0}=0$, $r_{m}=1$ and $r_{i}<r_{j}$ for $i<j$. $N_{err}$ is the number of times that the dimer constraint is violated in the swapped
configuration\cite{error}. If the difference between the successive $r_{i}$ is chosen to be sufficiently small, then each ratio in the product can be simulated
efficiently with Monte Carlo sampling. The essence of the ratio trick is to find a smooth interpolation between the distribution in the numerator of Eq.(2) and that in the denominator of Eq. (2), which are realized when $r_{0}=0$ and $r_{m}=1$.

To extract the value of TEE from $S_{2}$, we adopt the Kitaev-Preskill subtraction scheme and divide
the subsystem into three smaller parts, A, B and C, as is illustrated in Figure 2. Both A and B contain
$n_{a}\times n_{a}$ sites and C contains $n_{a}\times 2n_{a}$ sites. Then we have
\begin{eqnarray}
-\gamma=S_{2}^{A}+S_{2}^{B}+S_{2}^{C}-S_{2}^{AB}-S_{2}^{AC}-S_{2}^{BC}+S_{2}^{ABC}.\nonumber
\end{eqnarray}
This algorithm has been used to demonstrate the $\mathrm{ln}2$ TEE for the triangular QDM\cite{Pei,Furukawa}.

 \begin{figure}[h!]
\includegraphics[width=8cm,angle=0]{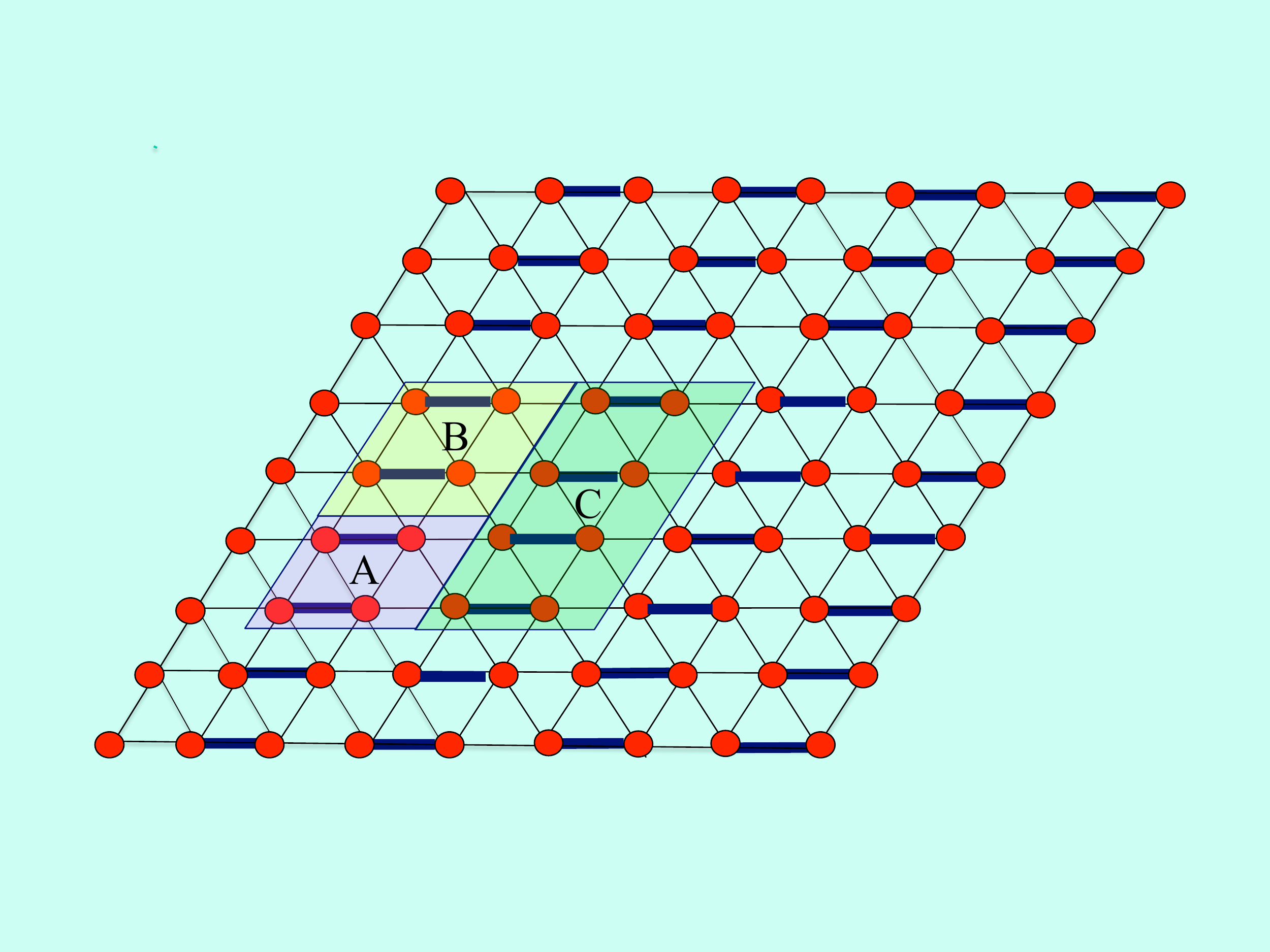}
\caption{The Kitaev-Preskill partition of the QDM used to calculate the TEE of the system. The subsystem is divided into three smaller parts $A$,$B$ and $C$ with linear size $n_{a}\times n_{a}$, $n_{a}\times n_{a}$ and $n_{a}\times 2n_{a}$. Illustrated here is the case for $n_{a}=2$. Note that the partition is defined for the bonds, rather than the sites of the lattice. We adopt the convention that a bond whose center belongs to a region is defined as a bond belonging to the same region. Region A and B thus have different number of bonds and are not equivalent under lattice translation. } \label{fig2}
\end{figure}

\begin{figure}[h!]
\includegraphics[width=9cm,angle=0]{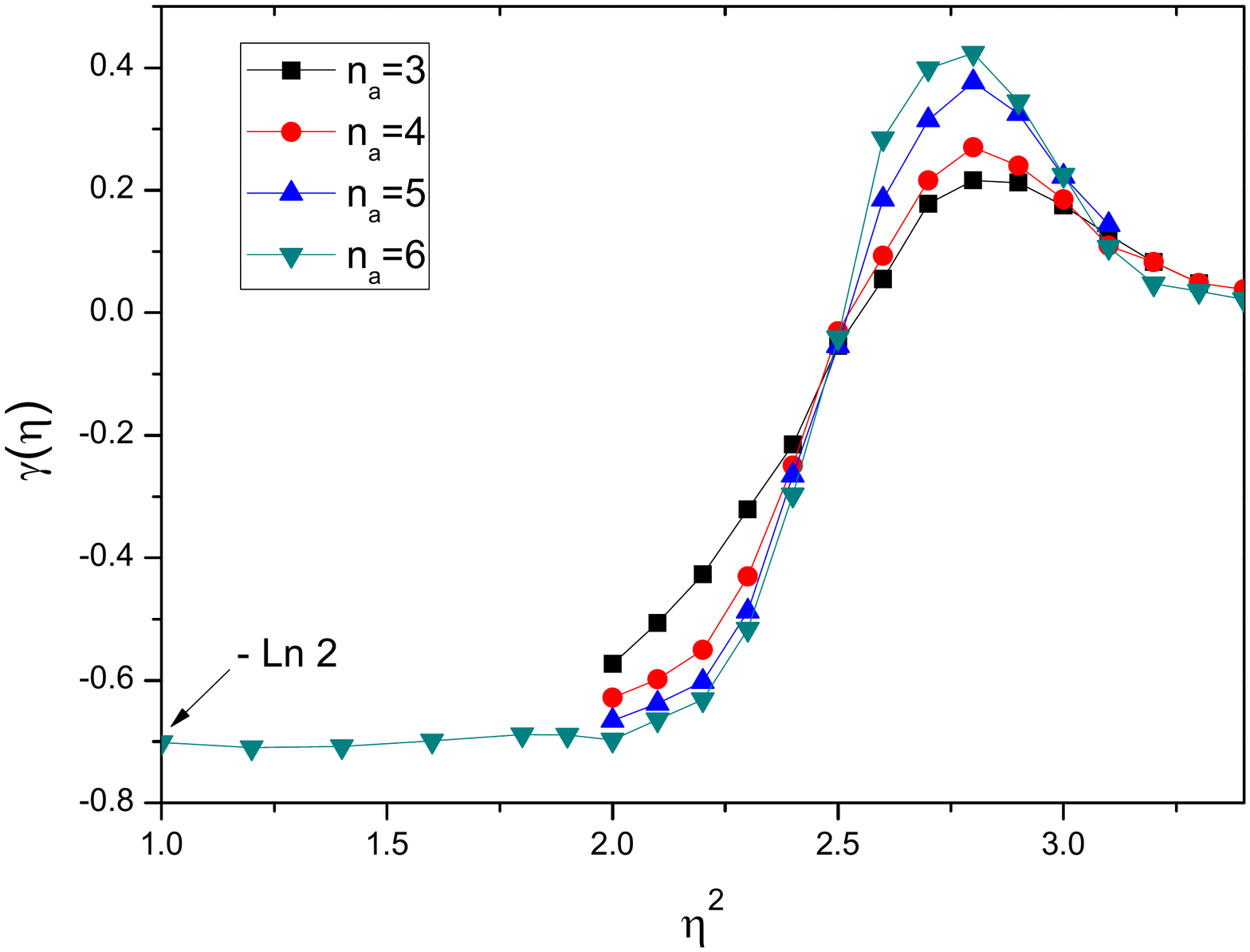}
\caption{Evolution of the TEE with the parameter $\eta$. The size of the torus is $24\times24$. $n_{a}$ is the linear size of the subsystem $A$. } \label{fig3}
\end{figure}

We have adopted the above algorithm to study the topological property of the generalized R-K state $|\Psi\rangle$. The system is defined on a $L\times L$ tours. Here we set $L=24$. The subsystem size considered varies from $n_{a}=3$ to $n_{a}=6$. We have introduced 10 intervals for $r_{i}$ in the ratio trick. For each ratio in the product, we have performed Monte Carlo simulation with up to $5\times10^{7}$ independent samples.  

The result of TEE as a function of $\eta$ is plotted in Figure 3. The most prominent feature in the plot is the broad peak in $\gamma(\eta)$ around $\eta^{2}_{c}\approx 2.75$. For $\eta \ll \eta_{c}$, $\gamma(\eta)$ converges to $\mathrm{ln}\ 2$, the value expected for a $Z_{2}$ topological ordered phase. For $\eta \gg \eta_{c}$, $\gamma(\eta)$ converges to zero, as expected for a trivial solid phase. The broad peak in $\gamma(\eta)$ is thus the signature of phase transition between a topological non-trivial solid phase and a trivial solid phase. 

In principle, both $L$ and $n_{a}$ should be extrapolated to infinity to extract the value of $\gamma(\eta)$. For the fixed torus size of $L=24$, we find that $\gamma(\eta)$ converges gradually to $\mathrm{ln}2$ when $\eta<\eta_{c}$ as we increase the subsystem size $n_{a}$, provided that $2n_{a}<L/2$\cite{finite}. We thus believe that the $Z_{2}$ topological order is robust in the whole region of $1<\eta<\eta_{c}$. For the same reason, we believe that the system is topological trivial when $\eta>\eta_{c}$. We thus expect that the width of the peak in $\gamma(\eta)$ should shrink to zero when both $L$ and $n_{a}$ are extrapolated to infinity. Calculation for larger $L$ and $n_{a}$ is definitely needed to confirm such an expectation. However, the calculation of TEE for large $L$ and $n_{a}$ becomes prohibitively expensive as a result of the area law in entanglement entropy, which dictates that the value of $\mathrm{Tr}\rho^{2}_{s}$ should decrease exponentially with the boundary length of the subsystem. At the same time, larger subsystem size is needed to extract the intrinsic value of $\gamma$ as we approach $\eta_{c}$, where the gapless vison fluctuation is expected to generate additional correction to the entanglement entropy. It is thus very hard to extend the scale of our calculation substantially.

The transition as revealed by $\gamma(\eta)$ is beyond the Landau's paradigm of phase transition, since the symmetry on both sides of the transition point are identical.  Actually, such a transition should be understood as a topological phase transition between a topological non-trivial solid phase and a trivial solid phase. The existence of such a non-Landau transition is direct proof for the coexistence between the $Z_{2}$ topological order and the columnar VBS order below $\eta_{c}$. It also implies that fractionalized excitation can coexist with conventional symmetry breaking order.

The origin of the transition in $\gamma(\eta)$ can be understood more clearly in terms of vison condensation\cite{Ivanov,Ralko2,Ralko3}. A vison is a $Z_{2}$ vortex in the QDM, around which the product of dimer amplitudes acquires an additional minus sign. The $Z_{2}$ topological order is characterized and protected by an nonzero vison excitation gap, while the trivial solid phase should support gapless vison excitation. To illustrate this point, we have calculated the vison correlation function in the generalized R-K state. The correlation function between two vison excitations can be found from the expectation value of the operator $(-1)^{n_{\Gamma}}$, in which $n_{\Gamma}$ is the number of the dimers that are intersected by the cut line $\Gamma$ connecting the two vison excitations(see Figure 1). Here we use the vison correlation function at the largest distance on the finite cluster as an estimate of the magnitude of the vison condensate. The result of the vison condensate defined in this manner is plotted in Figure 4, where the torus size is $L=36$. The critical behavior of the vison correlation seems to indicate a continuous transition between the topological solid phase and the trivial solid phase. A more though study is definitely needed to learn more accurately the critical behavior of this phase transition.

\begin{figure}[h!]
\includegraphics[width=8cm,angle=0]{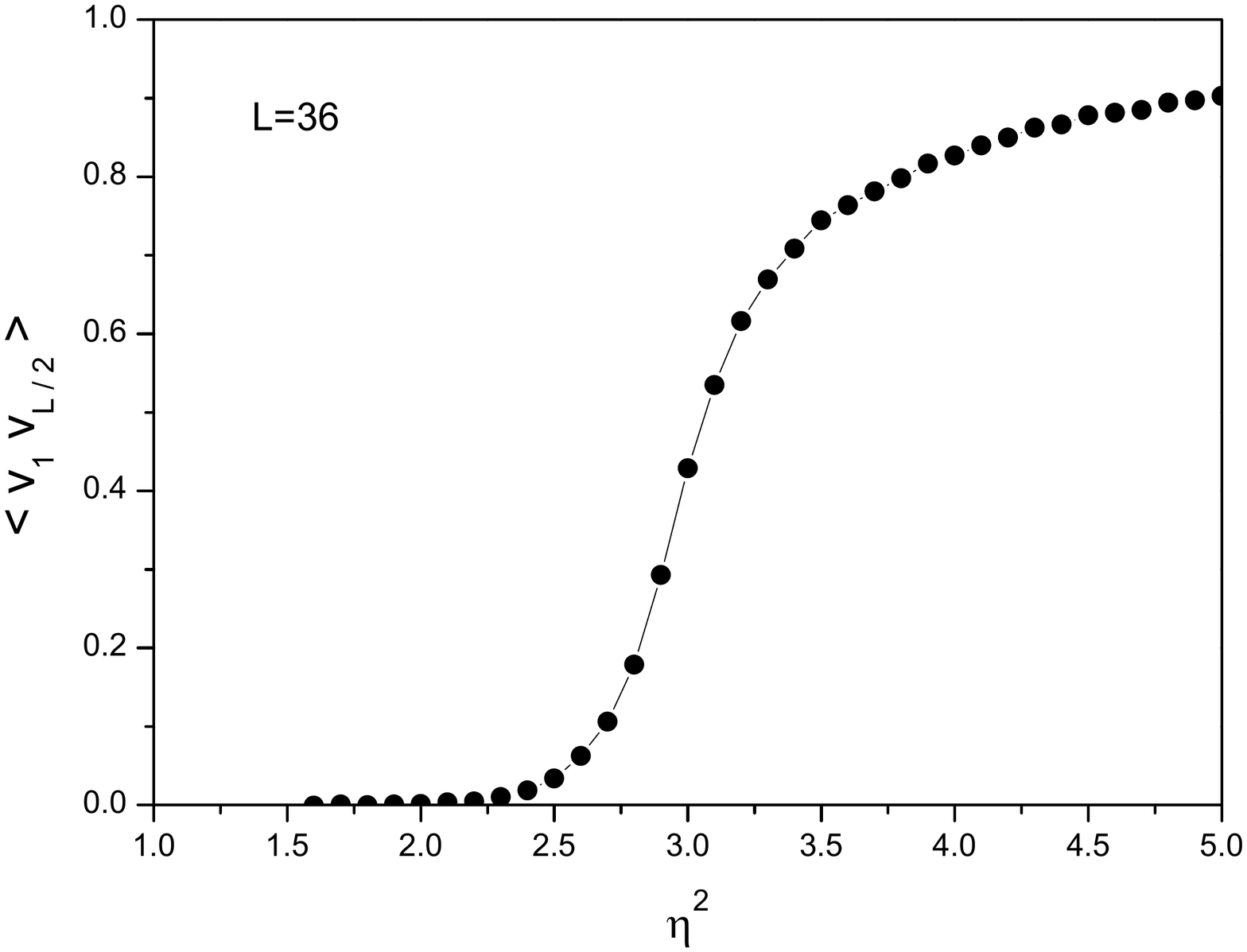}
\caption{The correlation function between two visons at the largest distance on a $36\times36$ torus in the $\vec{a}_{1}$ direction of the triangular lattice. $\vec{a}_{1}$ is the translation vector in direction of the thick bond in Figure 1.} \label{fig4}
\end{figure}

We note that the generalized R-K state proposed in this work has the same symmetry as the columnar VBS phase of the triangular QDM\cite{Sondhi,Ralko1}. It is however not clear if the columnar VBS phase realized by the standard triangular QDM is topological trivial or not. Subtle modification on the standard QDM may be necessary to stabilize the topological solid phase proposed in this paper. This is very similar to the situation of supersolid which, although allowed by physical principle, is highly non-trivial to find model realization.

In conclusion, through the calculation of the topological entanglement entropy of a generalized R-K state of the triangular QDM, we established a concrete example for the coexistence of the $Z_{2}$ topological order and conventional symmetry breaking order. Our result implies that the fractionalized excitation is not necessarily confined in the background of conventional symmetry breaking order. We believe that such coexistence is not limited to the lattice symmetry breaking studied in this work, but is relevant for general Landau symmetry breaking order. In particular, we expect that the topological order can coexist with the magnetic order. This may be relevant for the recent experimental observation of spinon-continuum-like spectral feature in several magnetic ordered quantum magnets\cite{AF1,AF2,AF3}.


\begin{thebibliography}{9}
\bibitem{error}An illustration of the exact meaning of $N_{err}$ can be found in Ref. [20].
\bibitem{finite}When $2n_{a}>L/2$, the complement of the region ABC on the torus will have a width that is smaller than that of the region ABC.
\end{thebibliography}
\end{document}